\documentclass[10pt,journal,compsoc]{llncs}

\usepackage{epsfig}
\usepackage{graphicx}
\usepackage{amsmath}
\usepackage{amssymb}
\usepackage{algorithm,algpseudocode}
\usepackage{color}
\usepackage{caption}
\usepackage{subcaption}
\usepackage{times,balance,multirow}
\usepackage{array}
\usepackage{adjustbox}
\usepackage{cite}
\usepackage{braket}
\usepackage{cases}
\usepackage{nicefrac}
\usepackage{booktabs}
\usepackage{soul}
\usepackage{makecell}
\usepackage{threeparttable}
\usepackage{arydshln}
\usepackage{mathtools}
\usepackage{hyperref}

\DeclareMathOperator*{\argmaxB}{argmax}   

\newcommand{\add}{\textcolor{black}}

\begin{document}

\title{PixelSteganalysis: Pixel-wise Hidden Information Removal with Low Visual Degradation}

\author{Dahuin Jung, Ho Bae, Hyun-Soo Choi,
       and~Sungroh Yoon*,~\IEEEmembership{Senior Member,~IEEE}
\IEEEcompsocitemizethanks{\IEEEcompsocthanksitem D. Jung, and S. Yoon are with the Dept. of Electrical and Computer Engineering, Seoul National University, Seoul 08826, Korea.
\IEEEcompsocthanksitem H. Bae is with Dept. of Cyber Security, Ewha Womans University, Seoul, 03760, Korea.
\IEEEcompsocthanksitem H.S. Choi is with Dept. of Computer Science \& Engineering and Interdisciplinary Graduate Program in Medical Bigdata Convergence in Kangwon National University, Chuncheon, 24341, Korea, and also with Ziovision, Chuncheon, 24341, Korea. 
\IEEEcompsocthanksitem S. Yoon is also with Interdisciplinary Program in Artificial Intelligence, Seoul National University, Seoul 08826, Korea
\IEEEcompsocthanksitem Correspondence should be addressed to S. Yoon.~(sryoon@snu.ac.kr).}}

\markboth{IEEE TRANSACTIONS ON Dependable and Secure Computing}
{Jung \MakeLowercase{\textit{et al.}}: PixelSteganalysis: Pixel-wise Hidden Information Removal with Low Visual Degradation}

\IEEEtitleabstractindextext{
\begin{abstract}
Recently, the field of steganography has experienced rapid developments based on deep learning (DL). DL based steganography distributes secret information over all the available bits of the cover image, thereby posing difficulties in using conventional steganalysis methods to detect, extract or remove hidden secret images. However, our proposed framework is the first to effectively disable covert communications and transactions that use DL based steganography. We propose a DL based steganalysis technique that effectively removes secret images by restoring the distribution of the original images. We formulate a problem and address it by exploiting sophisticated pixel distributions and an edge distribution of images by using a deep neural network.~Based on the given information, we remove the hidden secret information at the pixel level. We evaluate our technique by comparing it with conventional steganalysis methods using three public benchmarks. As the decoding method of DL based steganography is approximate (lossy) and is different from the decoding method of conventional steganography, we also introduce a new quantitative metric called the destruction rate (DT). The experimental results demonstrate performance improvements of 10--20$\%$ in both the decoded rate and the DT. 
\end{abstract}

}

\maketitle

\IEEEdisplaynontitleabstractindextext

%
\IEEEpeerreviewmaketitle

\IEEEraisesectionheading{\section{Introduction}\label{sec:introduction}}
\IEEEPARstart{S}{teganography} is the technique of unnoticeably concealing a secret message within a plain cover image to covertly send a message to an intended recipient~\cite{johnson1998exploring}. When a secret message is hidden in a cover image, the output is called a stego image. With the upsurge of big data on the Internet, the threat of unauthorized and unlimited information transaction and display has risen sharply. Furthermore, steganography has been used by international terrorist organizations, various companies, and military organizations for covert communications~\cite{sharma2012comparative}. It is known that some terrorist groups use steganography to exchange secret messages~\cite{gardner_2013}; it is also being used to steal confidential information from companies~\cite{king2018}. 

In the process of covertly embedding a secret message into a cover image, the original cover image is marginally altered to become a stego image~\cite{johnson1998exploring,lie1999data}. In conventional steganography, the payload of the secret message is small, and secret messages are mostly embedded in the least significant bits (LSBs) of the cover image in order to avoid statistical and visual detection~\cite{pevny2010using,holub2012designing}. Hence, the secret messages hidden using conventional steganography could be removed via relatively simple steganalysis techniques such as JPEG compression~\cite{fridrich2002steganalysis}. The decoding method of conventional steganography is lossless, therefore it can well retain the content of the hidden text.

As deep learning (DL) techniques have demonstrated great performance in various fields, so do steganography techniques exploiting DL techniques~\cite{dong2018invisible,wu2018stegnet}. The currently proposed DL based steganography disperses the representations of secret images across all the available bits~\cite{baluja2017hiding} and is not restricted to LSBs. The payload of a secret message embedded using a DL based steganography method is comparatively large; however, because the decoding method of DL based steganography is approximate (lossy), secret messages are mostly limited to the image form.

Steganalysis is the detection, extraction, or destruction of a secret message hidden in a stego image~\cite{johnson1998steganalysis,bachrach2011image}. Depending upon the privilege levels, steganalysis can be categorized as passive or active~\cite{amritha2016removal}. Passive steganalysis algorithms aim to determine whether an image contains a secret message. Most passive steganalysis algorithms look for features associated with a particular steganography technique (i.e., non-blind technique). However, active steganalysis algorithms involve blind techniques having the privilege to modify images. Active steganalysis represents the techniques used to extract and/or remove secret messages. However, most active steganalysis approaches try to remove the secret messages because extracting the exact messages is difficult in general cases~\cite{karaman2012application}. The original images processed after using active steganalysis must be nearly as unchanged as possible because not all images are stego images. That is, a good active steganalysis technique should aim to remove a secret message as much as possible while introducing minimal changes to the appearance of the image. To that end, we propose a new method of removing the secret image by restoring the distribution of the stego image to that of the original cover image. In Fig.~\ref{fig:activesteganalysis}, we illustrate an overview of steganography and active steganalysis with symbols of each process and material. The contributions of this paper are as follows:

\begin{figure*}[t]
\centering
\includegraphics[width=1.\linewidth]{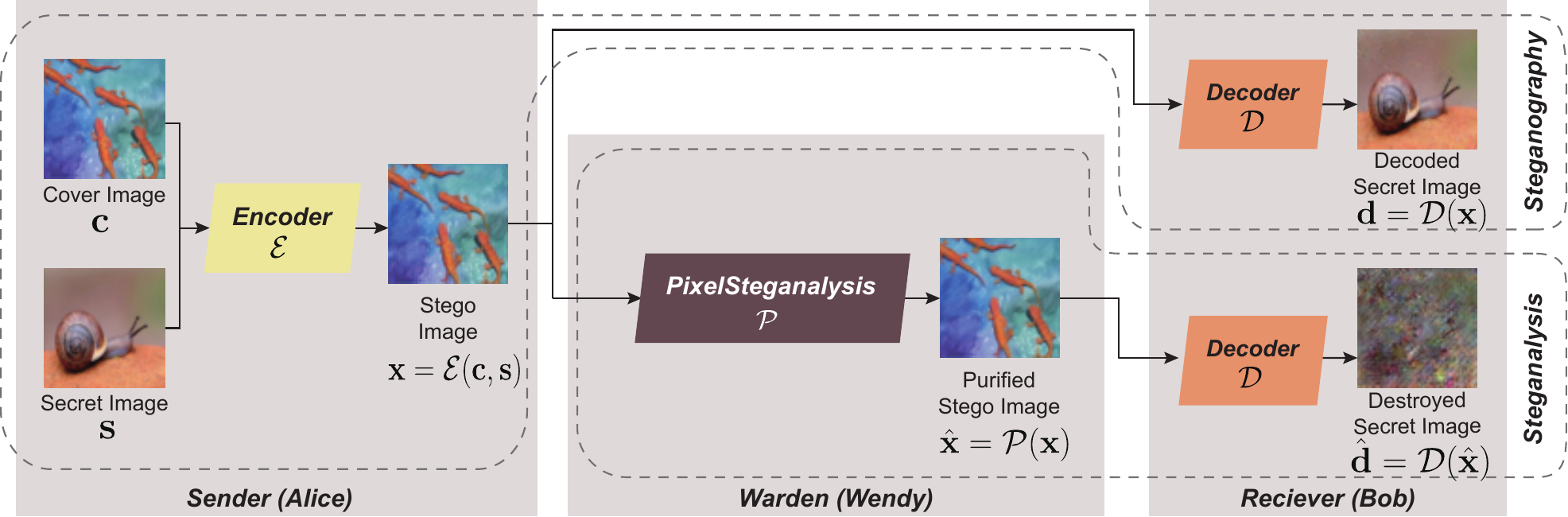}
\caption{How steganography works (Encoder $\mathcal{E}$ and Decoder $\mathcal{D}$) and how active steganalysis can disturb it. The destroyed secret image $\hat{\mathbf{d}}$ shows that PixelSteganalysis $\mathcal{P}$ disrupts a covert transmission between a sender and a receiver with an imperceptible difference on the stego image.}
\label{fig:activesteganalysis}
\end{figure*}

\begin{itemize}
\item To the best of our knowledge, this is the first method that effectively removes secret hidden images using a DL based steganography method. This is the first study utilizing DL to restore stego images to the original cover images. We present a theoretical formulation of the problem and its objectives. Furthermore, we experimentally show the possibility of using our approach as a passive steganalysis technique.

\item We assume a real-world situation in which access to either the cover image or the secret image is not allowed. Our framework only utilizes a dataset commonly associated with the target society (neither the cover image nor the secret image itself).

\item We propose a new evaluation metric called the destruction rate (DT) suitable for evaluating the performance of active steganalysis against the DL based steganography methods with lossy characteristics.

\item Our method outperforms conventional active steganalysis from both DL based steganography and conventional steganography, with both high and low payloads. Compared with the adaptive Gaussian noise method, our method exhibited improvements of up to 18$\%$ and 20$\%$ in terms of the peak signal-to-noise ratio (PSNR) and DT, respectively.
\end{itemize}

\add{The remainder of this paper is organized as follows.~\hyperref[sec:background]{Section 2} provides a brief description of related and comparison works.~\hyperref[sec:method]{Section 3} provides detailed descriptions of the attack scenario, problem formulation, and proposed methodology.~\hyperref[sec:metric]{Section 4} suggests a new evaluation metric, DT, that complements the limitation of the conventional evaluation metric.~\hyperref[sec:experiments]{Section 5} demonstrates the proposed methodology through various combinations of experiments. Finally,~\hyperref[sec:conclusions]{Section 6} discusses the results and areas of future study.}

\section{Background}
\label{sec:background}

\subsection{Conventional Active Steganalysis}
The destruction of the secret message hidden using conventional steganography was straightforward. Thus, active steganalysis has not been developed after several simple yet effective steganalysis approaches were proposed. The most basic approach is to take $N$ LSB planes of the stego image and flip the bits~\cite{ettinger1998steganalysis}. Another commonly used strategy for destructing a secret message is to overwrite the LSB bits randomly using Gaussian noise or other noise~\cite{fridrich2002steganalysis}. Although applying adaptive randomization into LSB bits can have comparably high destruction capability on spatial steganography algorithms, yet, at the same time, it can largely harm the image quality. Furthermore, \add{filter-based constructive destruction approaches have been proposed~\cite{gou2007noise,amritha2016removal,shrestha2011general,lafferty2008obfuscation}.} First, denoising is used to remove the hidden secret messages, considering the secret message as a noise added to the cover image ~\cite{gou2007noise}. In addition, Wiener restoration is a representative method for conventional active steganalysis~\cite{amritha2016removal}. The median filter (a denoising technique) and Wiener restoration both operate quite optimally on frequency domain steganography algorithms. We compared adaptive randomization, denoising, and Wiener restoration methods with the steganalysis method proposed in this paper because these three methods have been demonstrated as effective on conventional steganography~\cite{lafferty2008obfuscation}.

\subsection{Conventional and DL based Steganography}
\add{Steganography, unlike watermarking~\cite{watermarking}, aims at covert transmission through invisible data hiding. For it, various methods have been proposed to increase both the hiding capability and invisibility.} However, the hiding capability and invisibility have a contradictory relation in the field of steganography~\cite{imperceptibility}. Because conventional steganography is aimed at perfect invisibility, an extremely small hiding capability is generally maintained. The LSB insertion~\cite{johnson1998exploring,mielikainen2006lsb} is the most conventional steganographic algorithm. However, it is statistically obvious. Recent studies have proposed more advanced approaches that design diverse distortion functions to maintain image statistics. The Highly Undetectable steGO (HUGO) method is the first steganography method that devises a distortion function~\cite{pevny2010using,filler2010gibbs}. The HUGO method embeds the secret information at the positions where the difference between the features of the cover and stego images in the SPAM feature space is low, such as at an edge. In the spatial domain, the Wavelet Obtained Weights (WOW) method computes the sums of the weighted changes in the horizontal, vertical, and diagonal wavelet coefficients. The WOW method, then, estimates a pixel whose probability of being revealed is high in at least one direction by using a reciprocal Holder norm~\cite{holub2012designing}. On the basis of the estimated pixels, the WOW method can avoid clean edge areas while embedding secret information. As a disadvantage of WOW, the embedding cost is not sufficiently sensitive to texture regions because of merely adding the reciprocal norm rather than relative changes to the wavelet coefficient~\cite{furthers-uni}. Furthermore, the Spatial - UNIversal WAvelet Relative Distortion (S-UNIWARD) method is similar to the WOW method. However, the S-UNIWARD method addressed and moderated the above mentioned disadvantages of the WOW method~\cite{holub2014universal}.

Several DL based steganography approaches attempted to hide secret information with a very small payload~\cite{hayes2017generating,shi2017ssgan,zhu2018hidden}. However, as DL developed, DL based steganography began to embed secret messages of bigger payloads, such as a full-size image, into the cover image, with improved capacity~\cite{wang2019stnet,huang2019image,wang2019hidinggan,meng2019novel,chen2020high,kuppusamy2020novel,dlstega1,dlstega2,dlstega3,dlstega4,wu2018stegnet,yu2018integrated,baluja2017hiding,dong2018invisible, udh}. DL based steganography focuses on hiding a much larger amount of secret information in the cover image by relaxing the constraint of perfect invisibility. StegNet~\cite{wu2018stegnet} proposed an additional loss term named variance loss, which can reduce the noisiness of a stego image produced by generator networks. ISS-GAN~\cite{yu2018integrated} introduced a cycle discriminative structure and the concept of inconsistent loss, both of which can improve the quality and security of a stego image. Deep Steganography (DS)~\cite{baluja2017hiding} involves the additional use of a prep-network to transform a color-based secret image to an edge-based secret image for more natural and compact embedding. ISGAN~\cite{dong2018invisible} uses only the Y component from the YCbCr color space of the cover image to hide secret gray images, so that the destruction of the hidden secret images is more difficult than that using other methods. 

\begin{figure}[t]
\centering
\includegraphics[width=1\linewidth]{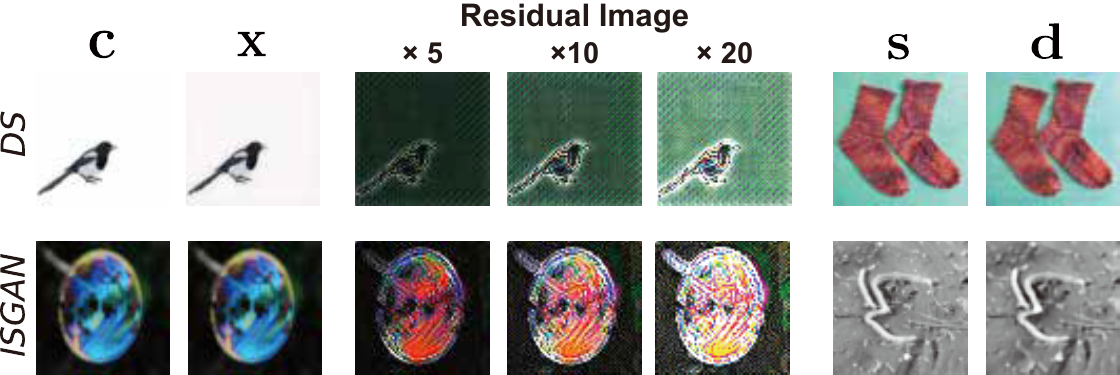}

\caption{Samples of DS~\cite{baluja2017hiding} and ISGAN~\cite{dong2018invisible}. The labels x5, x10, and x20 represent the magnification ratio (five, ten, and 20 times, respectively) of the residual image. The first row shows the residual outputs between the cover image $\mathbf{c}$ and the stego image $\mathbf{x}$ generated by DS. We can observe that the diagonal grid pattern is distributed all over the background of the residual images. The second row shows the residual outputs between the cover image $\mathbf{c}$ and the stego image $\mathbf{x}$ generated by ISGAN. For ISGAN, the stego image $\mathbf{x}$ looks natural despite the large difference between the cover image $\mathbf{c}$ and stego image $\mathbf{x}$. However, the illumination of the decoded secret image $\mathbf{d}$ is noticeably deviated from the secret image $\mathbf{s}$.}
\label{fig:edge_residual_example}
\end{figure}

The DL based steganography approaches cannot be detected easily using conventional passive steganalysis because these approaches tend to maintain the pixel distributions of the original image to the greatest extent possible~\cite{hayes2017generating,dong2018invisible}, which was demonstrated experimentally by Baluja and Shumeet~\cite{baluja2017hiding}. We also test whether conventional and DL based passive steganalysis methods can detect stego images created by DL based steganography methods. For testing, we use two representative statistical passive steganalysis techniques, RS~\cite{chaum1988dining} and SPA~\cite{berthold2001web}, and one DL based passive steganalysis technique, YeNet~\cite{ye2017deep}. We confirm experimentally that RS, SPA, and YeNet were unable to determine the stego images created by DL based steganography methods when assuming real-world situations in which no access to the utilized steganography algorithm is allowed.

Yu and Chong~\cite{yu2018integrated} demonstrated that secret images hidden using DL based steganography cannot be removed using conventional active steganalysis. Similar to conventional steganography methods, DL based steganography methods also hide the majority of secret information in the high-frequency areas of the cover image, such as edge, where the bandwidth is sufficiently wide to hide a considerable amount of information naturally. \add{It is also known from information theory that a change in the high-frequency area is hard to discover~\cite{informationtheory}.} More specifically, in the case of DS, 48.32$\%$ of the secret image is hidden in the edge areas on average, notwithstanding the fact that the edge area is only a small part of the entire image (approximately only 20$\%$ in a natural image). Moreover, in the case of ISGAN, 34.47$\%$ of the secret image is hidden in the edge areas on average. UDH~\cite{udh} recently proposed a meta-architecture that can disentangle the encoding of the secret image from the cover image. They conducted further analysis about where and how the secret image is encoded. A visual sample in which secret information is heavily hidden in the edge areas is provided in Fig.~\ref{fig:edge_residual_example}. Based on the analysis results, we propose a method that can remove secret information hidden using DL based steganography as much as possible while reducing the loss of the original cover image. In addition, we experimentally confirm the effectiveness of our approach on conventional steganography.

\begin{figure}[t]
\centering
\includegraphics[width=1.0\linewidth]{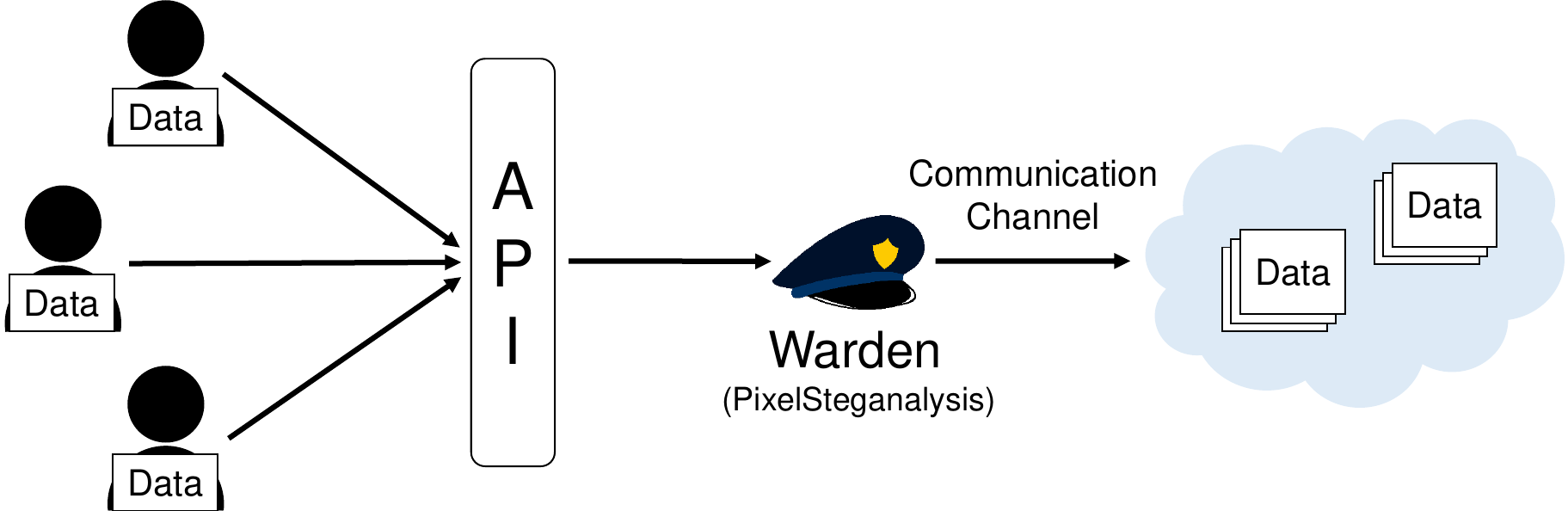}
\caption{\add{The attack scenario of the covert transmission of restricted information, based on Simmons' Prisoners' Problem~\cite{simmons1984prisoners}.}}
\label{fig:scenario}
\end{figure}

\begin{figure*}[t]
\centering
\includegraphics[width=1\textwidth]{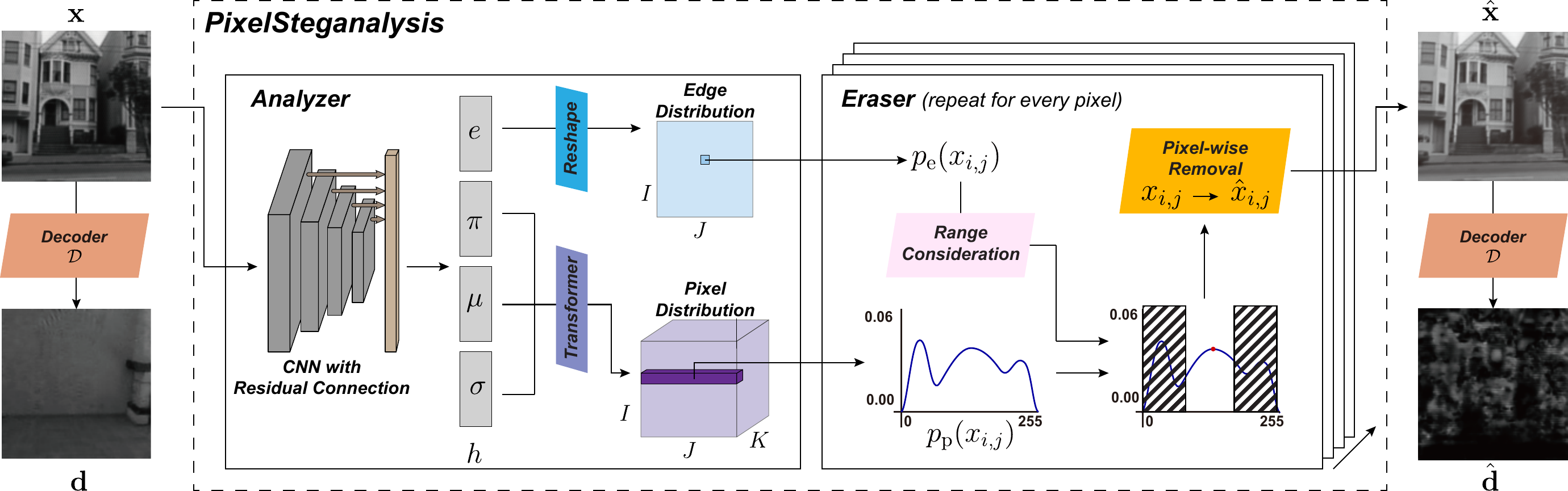}
\caption{Model overview. Our framework, PixelSteganalysis $\mathcal{P}$, consists of the $\textit{analyzer}$ and $\textit{eraser}$. PixelSteganalysis $\mathcal{P}$ receives the stego image $\mathbf{x}$ and produces the purified stego image $\hat{\mathbf{x}}$ ($\hat{\mathbf{x}} = \mathcal{P}(\mathbf{x})$).}
\label{fig:overview}
\end{figure*}

\section{Proposed Method}
\label{sec:method}

\subsection{Scenario} \label{sec:scenario}

\add{There is always a risk that individuals having access to sensitive or proprietary information try to leak information and then share it with competitors or adversaries. We particularly assume a scenario that the local hosts in restricted environments such as companies inside try to leak hidden information via the internet. As depicted in Fig.~\ref{fig:scenario}, our framework can be located in the company’s uplink to the internet outside as an active warden~\cite{simmons1984prisoners}.}

\add{An active warden has the privilege to alter the content of the communication to confuse hidden data within the carrier. However, its privilege is limited to slight changes~\cite{lafferty2008obfuscation}. As an active warden, our method aims at removing the hidden secret message in a direction of restoring the distribution of the original images.}

\subsection{Problem Formulation}

We represent the stego image, original cover image, secret image, and decoded secret image as $\mathbf{x}$, $\mathbf{c}$, $\mathbf{s}$, and $\mathbf{d}$, respectively. Then, we can formulate the encoding and decoding algorithms of steganography as $\mathbf{x} = \mathcal{E}(\mathbf{c}, \mathbf{s})$ and $\mathbf{d} = \mathcal{D}(\mathbf{x})$, respectively. We also represent a purified stego image after performing active steganalysis as $\hat{\mathbf{x}}$, and a destroyed secret image after performing active steganalysis as $\hat{\mathbf{d}} = \mathcal{D}(\hat{\mathbf{x}})$. We first formulate the objective of this study by using these representations. From the perspective of active steganalysis, the ultimate goal is to find the purified stego image $\hat{\mathbf{x}}$ as follows:
\begin{equation}
      \begin{aligned}
  & \underset{\hat{\mathbf{x}}}{\text{minimize}} & & \left \| \hat{\mathbf{x}} - \mathbf{c} \right \|_\infty.
   \label{eq:optimal_goal}
   \end{aligned}
\end{equation}
If $\mathbf{x}$ returns to $\mathbf{c}$, it indicates that the hidden secret image is removed completely after performing active steganalysis. In other words, the hidden secret image is completely extracted from the stego image $\mathbf{x}$. However, since the original cover image $\mathbf{c}$ and secret image $\mathbf{s}$ are inaccessible in the process of detecting or extracting the hidden secret image, it is almost impossible. As described in Sec.~\ref{sec:scenario}, rather than an utterly open scenario, we assume the scenario that works as an active warden in the restricted environment. Each company can accumulate and make use of the data samples transmitted outside from their uplink to build a dataset $\mathbf{X}$ having a similar distribution to the original cover image $\mathbf{c}$.
With these assumptions, we can alter Eq.~\ref{eq:optimal_goal} to 
\begin{equation}
      \begin{aligned}
  & \underset{\hat{\mathbf{x}}}{\text{maximize}} & & p^*(\hat{\mathbf{x}}) \\
  & \text{subject to} & & \left \| \hat{\mathbf{x}} - \mathbf{x} \right \|_\infty \leq \epsilon_{\mathrm{max}},
   \label{eq:relax_goal}
   \end{aligned}
\end{equation}
where $p^*(\cdot)$ denotes the exact distribution of the training dataset $\mathbf{X}$ and $\epsilon_{\mathrm{max}}$ denotes the allowed maximum modification of each pixel. That is, we aim to find an image $\hat{\mathbf{x}}$ that maximizes $p^*(\hat{\mathbf{x}})$ subject to the constraint that $\hat{\mathbf{x}}$ is within the $\epsilon_{\mathrm{max}}$-ball of $\mathbf{x}$. We constrain the maximum pixel modification because most steganography methods aim at a minimum change in $\mathbf{x}$ from $\mathbf{c}$, both statistically and visually. In practice, we propose an adaptive consideration degree of modification, $e_{\mathrm{norm}}$, bounded by $\mathrm{\epsilon_{\mathrm{max}}}$ as the constraint per pixel. This adaptive consideration range allows more changes in the edge areas and less change in the non-edge areas. Detailed descriptions are given in Eqs.~\ref{eq:eachenorm} and~\ref{eq:range}. We also approximate $p^*$ to a PixelCNN++ distribution $p$ (Eq.~\ref{eq:auto-regressive}). To maximally utilize the intrinsic characteristics of steganography, we propose a technique that sequentially satisfies the constrained objective in Eq.~\ref{eq:relax_goal} at the pixel level, instead of using gradient-based constrained optimization such as L-BFGS-B~\cite{zhu1997algorithm} or image generation~\cite{goodfellow2014generative}. 

\subsection{Proposed Algorithm}

To remove a hidden image, our proposed algorithm requires neither the knowledge of the utilized steganography algorithm (blind) nor the distribution of the original cover image, while offering minimum perceptual degradation and even perceptual improvement of the stego image. As illustrated in Fig.~\ref{fig:overview}, our algorithm employs an $\textit{analyzer}$ and an $\textit{eraser}$ to produce the purified stego image, $\hat{\mathbf{x}}$. The $\textit{analyzer}$ takes the stego image as input and produces an edge distribution, $\textbf{p}_{\mathrm{e}}( \mathbf{x})$, and the distribution of all the pixels, $\textbf{p}_{\mathrm{p}}( \mathbf{x})$, of the given image (Sec.~$\ref{ssec:sec331}$ \textit{Analyzer}). The generated distributions are then used to remove the secret image hidden in the stego image by using the $\textit{eraser}$ (Sec.~$\ref{ssec:sec332}$ \textit{Eraser}). Note that the candidate input images are not limited to grayscale. However, in this section, for easier visualization and explanation, we assume grayscale input images.

\subsubsection{\textit{Analyzer}} \label{ssec:sec331}

The $\textit{analyzer}$ obtains $\textbf{p}_{\mathrm{p}}( \mathbf{x})$ and $\textbf{p}_{\mathrm{e}}( \mathbf{x})$ by employing a neural network trained using a dataset $\mathbf{X}$ having similar distribution as the original cover images. \add{The auto-regressive models learn the image distribution. Then, we calculate the marginal likelihood of an image by taking the product of the probabilities of each sampled pixel}:
\begin{equation}
      \begin{aligned}
      & p(\mathbf{x}) =\prod_{i = 2}^{I \times J} p\left ( \mathbf{x}(i) |\mathbf{x}(1:(i-1)) \right ),
   \label{eq:auto-regressive}
   \end{aligned}
\end{equation}
where $I\times J$ denotes the height and width of the image. Therefore, we can take advantage of the explicit distribution of all the pixels, unlike other modeling algorithms. Moreover, $\textbf{p}_{\mathrm{e}}( \mathbf{x})$ is the information indicating the high frequency areas of the image. $\textbf{p}_{\mathrm{e}}( \mathbf{x})$ is jointly learned and is utilized in the $\textit{eraser}$. To obtain $\textbf{p}_{\mathrm{p}}( \mathbf{x})$ and $\textbf{p}_{\mathrm{e}}( \mathbf{x})$, we propose a CNN architecture inspired by PixelCNN++~\cite{salimans2017pixelcnn++}, which is the most representative DL based auto-regressive model.

As described in Fig.~\ref{fig:overview}, the activation of the last fully connected layer of the $\textit{analyzer}$ is named $\textit{h}$ and consists of trained parameters $\mathrm{e}, \pi, \mu,$ and $\sigma$. Using the parameters $\pi, \mu, \mathrm{and}$ $\sigma$ trained with a dataset $\mathbf{X}$ that has a distribution similar to the original cover image, we obtain a discretized Gaussian mixture likelihood for all pixels $\textbf{p}_{\mathrm{p}}( \mathbf{x}| \pi, \mu, \sigma)$ ($\in \mathbb{R}^{I\times J\times K}$) obtained by learning the distribution of the dataset $\mathbf{X}$, $p(\mathbf{\cdot})$, in an auto-regressive way, where $K$ denotes a pixel depth dimension (0--255). This procedure is referred to as the $\textit{transformer}$. Using the trained $\textbf{p}_{\mathrm{p}}$, we can obtain how appropriate the current pixel value is, provided the previous pixel values with respect to the distribution of cover images in the shape of a Gaussian mixture model. The operation of the $\textit{transformer}$ is based on PixelCNN++. The detail of the $\textit{transformer}$ can be found in supplementary S2 and~\cite{salimans2017pixelcnn++}.

\add{We also train a network to detect high frequency areas in which one can embed secret information unnoticeably, which we call an edge distribution, $\textbf{p}_{\mathrm{e}}( \mathbf{x})$.} 
As shown in Fig.~\ref{fig:overview}, $\textbf{p}_{\mathrm{e}}( \mathbf{x})$, the vector $\mathrm{e}$ of $\textit{h}$ is reshaped into $I\times J$. $\textbf{p}_{\mathrm{e}}( \mathbf{x})$ is used to determine the consideration range of the suspicious information per pixel in the $\textit{eraser}$, as explained in Sec.~\ref{sec:background}. 

We minimize the sum of the $\textit{image loss}$, $\mathcal{L}_I$, and the $\textit{edge loss}$, $\mathcal{L}_E$, to obtain $\textbf{p}_{\mathrm{p}}$ and $\textbf{p}_{\mathrm{e}}$, respectively. $\mathcal{L}_I$ is the negative log-likelihood of the image obtained by the product of the conditional distribution of each pixel, and $\mathcal{L}_E$ is the mean-squared error between the results obtained using a conventional edge detector and those obtained using our neural network. We have
\begin{equation}
    \begin{aligned}
      & \mathcal{L}_I = -\mathbb{E}_{\mathbf{x} \sim \mathbf{X}}\mathrm{log}\: p(\mathbf{x}), \\
      & \mathcal{L}_E = \mathbb{E}_{\mathbf{x} \sim \mathbf{X}}[\frac{1}{I \times J}  \sum_{i=1}^{I} \sum_{j=1}^{J} ({\textbf{p}_{\mathrm{d}}(\mathbf{x})(i, j)}  - {\textbf{p}_{\mathrm{e}}(\mathbf{x})(i, j)})^2], \\
      & \mathcal{L} = \lambda_I \mathcal{L}_I + \lambda_E \mathcal{L}_E,
   \label{eq:pixelcnn_output}
   \end{aligned}
\end{equation}
where $\mathbf{x}$ denotes the image of the training dataset $\mathbf{x} \in \mathbf{X}$, ${\textbf{p}_{\mathrm{d}}}( \mathbf{x})$ the edge distribution obtained using a conventional edge detector~\cite{prewitt1970object} currently, and $\textbf{p}_{\mathrm{e}}( \mathbf{x})$ the learned edge distribution. For empirical risk minimization, we make use of empirical expectations of each loss. Moreover, $\lambda_I$ and $\lambda_E$ $> 0$ denote the hyperparameters used to balance the strength of both the loss terms.

\subsubsection{\textit{Eraser}} \label{ssec:sec332}

The best scenario from the perspective of steganalysis is that both the cover and stego images are accessible. Then, we can easily restore a stego image, $\mathbf{x}$, to a cover image, $\mathbf{c}$, using Eq.~\ref{eq:optimal_goal}. However, this is generally impractical. Thus, instead, we suggest an approach for removing the hidden secret image by adjusting the pixel value of the suspicious regions in which the secret image may be hidden, using the pixel level information. In the $\textit{eraser}$, we aim to find a purified stego image, $\hat{\mathbf{x}}$, that maximizes $p(\hat{\mathbf{x}})$ under the constraint given in Eq.~\ref{eq:relax_goal}. \add{We iteratively substitute the pixel $i$'s value with the neighboring pixel value of the highest probability based on information from pixels $i - 1, i - 2, ...$.} 

A large amount of secret information is hidden in the edge areas of the cover image. Therefore, we control the consideration range of each pixel, and the range is decided by two factors: $\epsilon$ and $\textbf{p}_{\mathrm{e}}(\mathbf{x})$. We calculate the adaptive consideration degree of modification per pixel as follows:
\begin{equation}
      \begin{aligned}
  e_{\mathrm{norm}}(i, j) = \: \epsilon +  \: \lceil \frac{{\textbf{p}_{\mathrm{e}}(\mathbf{x})(i, j)}}{{\textbf{p}_{\mathrm{e}}(\mathbf{x})}_{\mathrm{max}}}\times (\epsilon_{\mathrm{max}} - \epsilon) \rceil,
   \label{eq:eachenorm}
   \end{aligned}
\end{equation}
where $i$ and $j$ denote the pixel coordinates of the image, ${\textbf{p}_{\mathrm{e}}(\mathbf{x})}_{\mathrm{max}}$ the maximum edge value and $\mathrm{\epsilon}$ a hyperparameter representing an allowed degree of the least modification ($\mathrm{\epsilon} > $ 0). We make use of a ceiling, $\left \lceil \cdot  \right \rceil$, to keep $e_{\mathrm{norm}}$ as an integer. The hyperparameter $\epsilon$ is suggested for fair comparisons with other active steganalysis methods and to guarantee the removal of encoded secret information in non-edge areas. Eq.~\ref{eq:eachenorm} shows that $e_{\mathrm{norm}}$ is bounded by $\epsilon$ and $\mathrm{\epsilon_{\mathrm{max}}}$ ($ \epsilon \leq e_{\mathrm{norm}} \leq \mathrm{\epsilon_{\mathrm{max}}}$). Eq.~\ref{eq:eachenorm} guarantees the lower and upper bounds of the consideration range per pixel. $\mathrm{\epsilon_{\mathrm{max}}}$ is $2 \times \epsilon$ in our experiments. 

\begin{figure}[t]
\centering
\includegraphics[width=1.\linewidth]{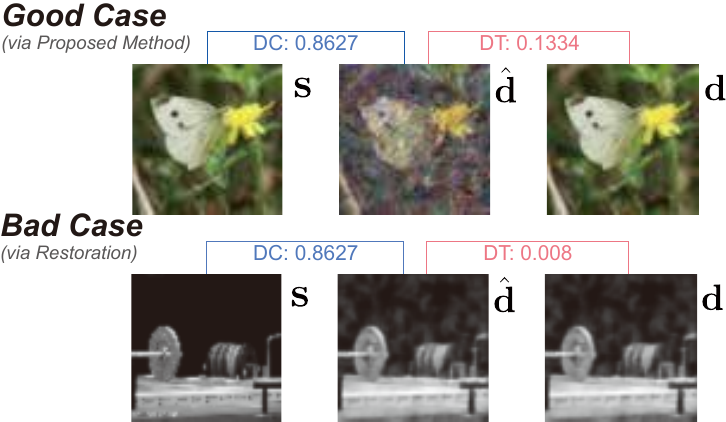}
\caption{Top: the success of destruction has a relatively high value of DT. Bottom: the failure of destruction is indicated by the almost zero DT values. However, the DC values of the two cases are exactly the same.}
\label{fig:badvsgood}
\end{figure} 

The pixel value of the stego image is not deviated significantly from the corresponding pixel value of the cover image. Therefore we only consider probabilities close to those of the given pixel values. We set up the adaptive consideration range of modification per pixel as: 
\begin{equation}
      \begin{aligned}
  & r_{\mathrm{min}}(i, j) = \max(\mathbf{x}(i, j) - e_{\mathrm{norm}}(i, j), \: 0), \\ & r_{\mathrm{max}}(i, j) = \min(\mathbf{x}(i, j) + e_{\mathrm{norm}}(i, j), \: 255).
   \label{eq:range}
   \end{aligned}
\end{equation} 
\add{Eq.~\ref{eq:range} determines the range of pixel values considered for modification by centering around $\mathbf{x}(i, j)$.} 

Each pixel is iteratively replaced with the one having the highest probability value among the allowed neighboring pixel values, as follows: 
\begin{equation}
      \begin{aligned}
  \hat{\mathbf{x}}(i, j) = \argmaxB_{k \in \left [ r_{\mathrm{min}}(i, j), r_{\mathrm{max}}(i, j)  \right ]} \textbf{p}_{\mathrm{p}}(\hat{\mathbf{x}})(i, j, k),
   \label{eq:argmax}
   \end{aligned}
\end{equation}
where initially $\hat{\mathbf{x}} = \mathbf{x}$. Eq.~\ref{eq:argmax} removes the secret message at the pixel level based on ${\textbf{p}_{\mathrm{p}}}(\mathbf{x})(i, j)$ bounded by $r_{\mathrm{min}}(i, j)$ and $r_{\mathrm{max}}(i, j)$. For every pixel, the pixel distribution $\textbf{p}_{\mathrm{p}}( \hat{\mathbf{x}})$ should be re-extracted whenever the previous pixel value is modified. However, re-extracting the pixel distribution for all the pixels requires excessive time to modify a single image. Therefore, to decrease the runtime, we propose an approximation of Eq.~\ref{eq:argmax}, in which the pixel distribution $\textbf{p}_{\mathrm{p}}( \mathbf{x})$ is only extracted before the iteration. The approximation is formed as
\begin{equation}
      \begin{aligned}
  {\hat{\mathbf{x}}}(i, j) = \argmaxB_{k \in \left [ r_{\mathrm{min}}(i, j), r_{\mathrm{max}}(i, j)  \right ]} \textbf{p}_{\mathrm{p}}(\mathbf{x})(i, j, k),
   \label{eq:argmax_relx}
   \end{aligned}
\end{equation}
for each pixel. This approximation leads to much faster runtime but slightly decreases the quality of the results. A comparison between the original modification case and the approximated modification case is presented in the Sec. \ref{ssec:sec52}.

\begin{figure*}[t]
\centering
\includegraphics[width=1.\textwidth]{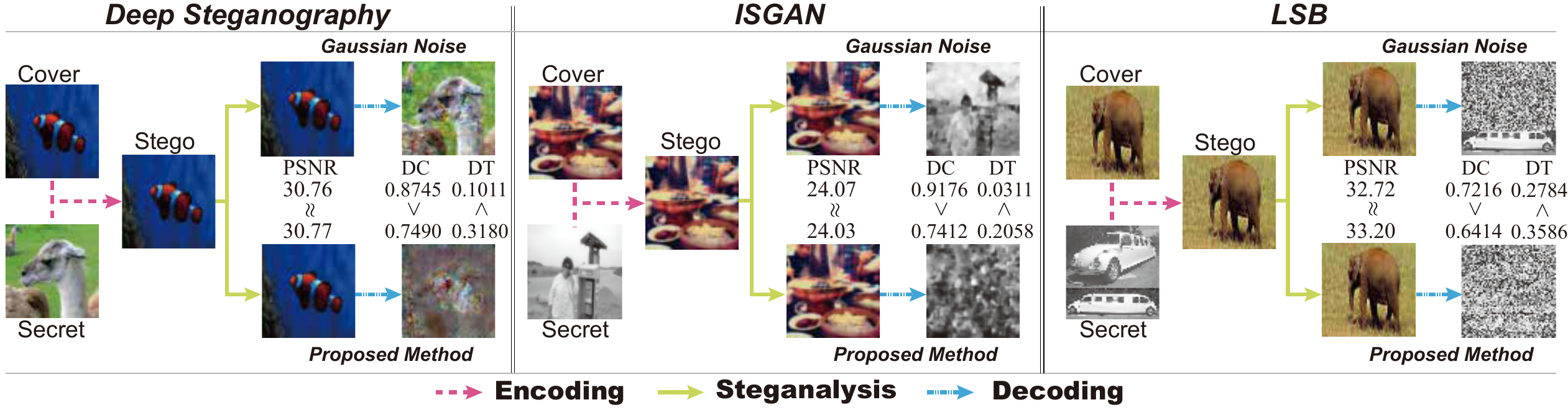}
\caption{Three examples of how our method and Gaussian noise differ in efficiency at the same PSNR.}
\label{fig:imagenet_total_output}
\end{figure*}

\section{Proposed Evaluation Metric}
\label{sec:metric}
The goal of an active steganalysis is the development of a technique that minimally destroys the cover image while effectively removing hidden steganography. To properly examine the performance of an active steganalysis method, an evaluation metric that defines the criteria to be met against various types of steganography approaches is necessary. However, the evaluation method against DL based steganography cannot separate the destruction of the hidden secret image from the degradation of the decoded secret image itself. In detail, the evaluation method against DL based steganography, the decoded rate (DC)~\cite{amritha2016removal,wu2018stegnet}, is defined by
\begin{equation}
      \begin{aligned}
  \mathrm{Decoded \: Rate} = 1 - \frac{\sum_{i=1}^{I} \sum_{j=1}^{J} \left | \mathbf{s}(i,j) - \hat{\mathbf{d}}(i,j) \right |}{I \times J}.
   \label{eq:decodedrate}
   \end{aligned}
\end{equation}
A condition that was guaranteed in conventional steganography is $\mathbf{s}$ = $\mathbf{d}$. Therefore, we could use the DC to measure the exact performance of active steganalysis. 
However, the decoding method of the DL based steganography algorithms is $\mathbf{s} \neq \mathbf{d}$ (lossy). The DC between $\mathbf{s}$ and $\mathbf{d}$ is approximately 90$\%$ \cite{baluja2017hiding,dong2018invisible}. In other words, the error rate between $\mathbf{s}$ and $\mathbf{d}$ is already 10$\%$, which depends upon how well the decoding algorithm is trained. Thus, we propose a new evaluation metric called the DT, which can accurately assess the destruction performance of active steganalysis on both conventional steganography and DL based steganography, regardless of the decoding algorithm. The DT is defined as
\begin{equation}
      \begin{aligned}
  \mathrm{Destruction \: Rate} = \frac{\sum_{i=1}^{I} \sum_{j=1}^{J} \left | \mathbf{d}(i,j) - \hat{\mathbf{d}}(i,j) \right |}{I \times J}.
   \label{eq:destructionrate}
   \end{aligned}
\end{equation}
To produce results independently of the performance of the decoding algorithm, the base image is changed into $\mathbf{d}$ instead of $\mathbf{s}$. DT is a more reliable metric than DC for representing the pure degree of hidden image destruction for each active steganalysis method. For example, as depicted in Fig.~\ref{fig:badvsgood}, it is possible that the DT values can be significantly different, whereas the DC values are exactly the same. Because the DC value is affected by the performance of both active steganalysis and the decoding method of steganography, its value can be meaningful even if the performance of the active steganalysis is poor.

\begin{figure}[t]
\centering
\includegraphics[width=1\linewidth]{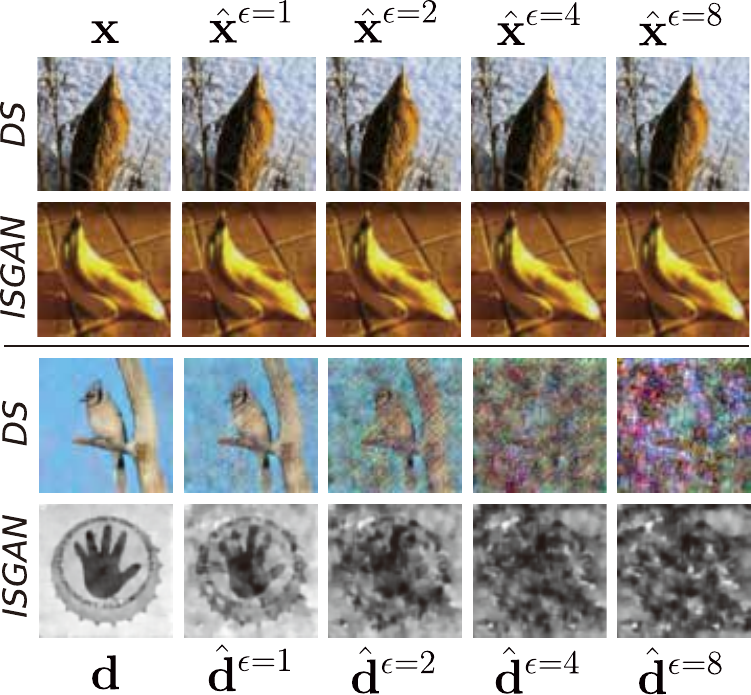}
\caption{According to the increase of $\epsilon$, the degree of destruction of the secret image hidden in the stego image (ImageNet) generated by DS and ISGAN after applying our method.}
\label{fig:change_by_epsilon}
\end{figure}

\section{Experiments}
\label{sec:experiments}

\subsection{Experimental Results}

We compare our method with three commonly used conventional steganalysis techniques: Gaussian noise, Denoising, and Restoration. We use the PSNR~\cite{psnr} and the structural similarity (SSIM)~\cite{ssim} to measure the quality of the purified image. PSNR and SSIM are basic metrics for comparing the quality of the purified image to that of the original cover image. The SSIM results are provided in supplementary S5. 

Among the DL based steganography algorithms, we made use of two representative methods, DS, ISGAN and UDH, to compare our method with the existing active steganalysis techniques. Additionally, we test our proposed steganalysis method on non-DL steganography techniques, namely, LSB insertion, HUGO, WOW, and S-UNIWARD. 

We made use of three datasets: Cifar-10, Boss1.0.1, and ImageNet. The results for the three datasets show similar trends. The details of the settings and the comparison methods are described in supplementary S1 and S6. 

\begin{figure*}[t]
\centering
\includegraphics[width=.92\linewidth]{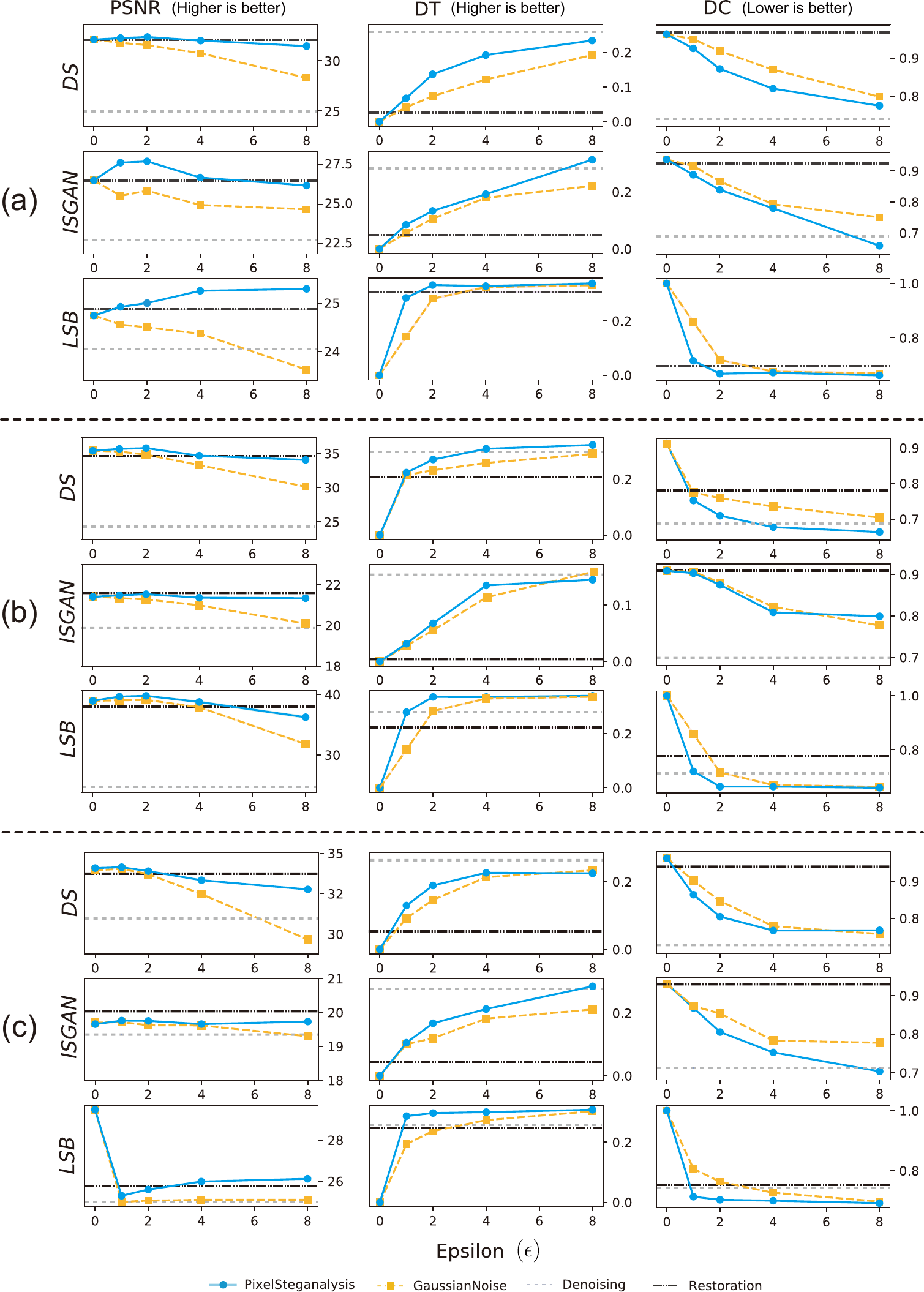}
\caption{Experimental results of our work and other steganalysis methods on the (a) Cifar-10, (b) ImageNet, and (c) Boss1.0.1 datasets. The higher the PSNR is, the better the preservation of the original cover image is. The lower the DC is and the higher the DT is, the better the destruction of the hidden secret image is.}
\label{fig:result_graph}
\end{figure*}

As shown in Fig.~\ref{fig:imagenet_total_output}, our method removes the hidden secret images better than the Gaussian noise method at the same PSNR in the three cases of steganography (DS, ISGAN, and LSB insertion). As we can observe, the DT value demonstrates the performance of active steganalysis more accurately. We also plot quantitative results for all the three datasets in Fig.~\ref{fig:result_graph}. In Fig.~\ref{fig:result_graph}, the PSNR at $\epsilon = 0$ represents the quality of the unpurified stego image compared with the cover image. Unlike the Gaussian noise method, the PSNR of our method did not fall below a certain level, even at $\epsilon = 8$ for all datasets. The reason is that we closely follow the distribution of the original cover image, so that regardless of how large $\epsilon$ is, the distribution of the purified stego image would not be substantially different from that of the cover image. 
\clearpage
\begin{figure*}[t]
\centering
\includegraphics[width=1\linewidth]{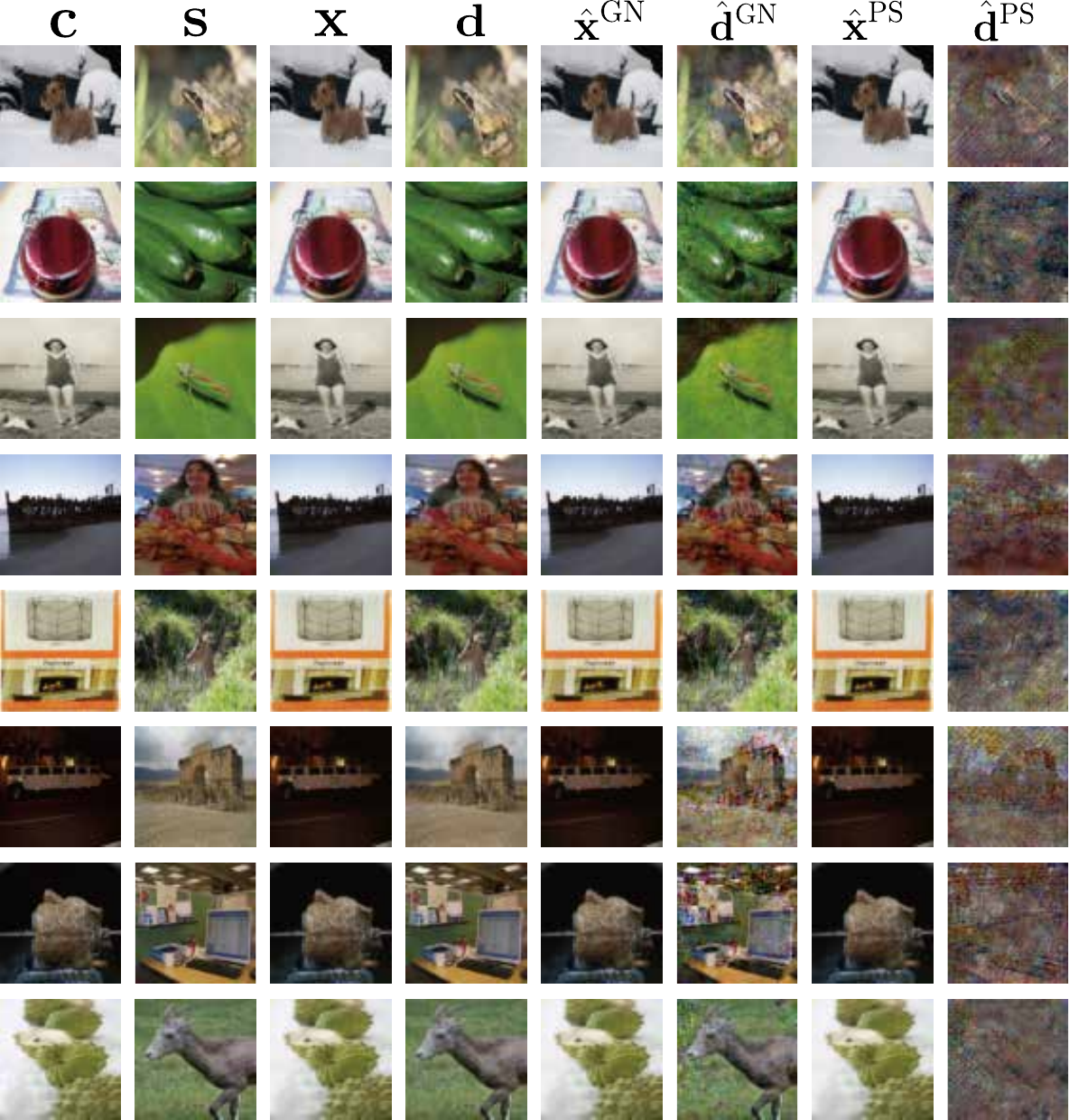}
\caption{Comparison of the destructed degree of the secret image decoded by the stego image from our method with that of the secret image decoded by the stego image from Gaussian noise  $\textbf{(DS on ImageNet)}$. $\hat{\mathbf{x}}^{\text{GN}}$ represents stego image modified by Gaussian noise, $\hat{\mathbf{d}}^{\text{GN}}$ represents secret image decoded from $\hat{\mathbf{x}}^{\text{GN}}$, $\hat{\mathbf{x}}^{\text{PS}}$ represents stego image modified by our method, and $\hat{\mathbf{d}}^{\text{PS}}$ represents secret image decoded from $\hat{\mathbf{x}}^{\text{PS}}$.}
\label{fig:dsimagenetpsnr}
\end{figure*}
\clearpage

\begin{table}[t]
\centering
\caption{PSNR (higher is better) and DC (lower is better) against HUGO($H$), WOW($W$), and S-UNIWARD($S$)}
\begin{adjustbox}{width=1.0\linewidth}
\begin{tabular}{ccccccccccccc}
\toprule
                & \multicolumn{6}{c}{IMAGENET (PSNR~\cite{psnr})}                                          & \multicolumn{6}{c}{IMAGENET (DC~\cite{amritha2016removal})}                           \\ \midrule
$\epsilon$      & \multicolumn{3}{c}{2}                & \multicolumn{3}{c}{4}                           & \multicolumn{3}{c}{2}                         & \multicolumn{3}{c}{4}                             \\ \cmidrule(l){1-1} \cmidrule(l){2-4} \cmidrule(l){5-7}  \cmidrule(l){8-10} \cmidrule(l){11-13} 
\textit{non-DL} & \textit{H} & \textit{W} & \textit{S} & \textit{H} & \textit{W} & \textit{S}  & \textit{H} & \textit{W} & \textit{S} & \textit{H} & \textit{W} & \textit{S} \\
\cmidrule(l){1-1} \cmidrule(l){2-4} \cmidrule(l){5-7}  \cmidrule(l){8-10} \cmidrule(l){11-13} 
\textbf{Proposed}        & \textbf{44.7}       & \textbf{44.7}       & \textbf{44.7}       & \textbf{40.3}       & \textbf{40.3}       & \textbf{40.3}       & \textbf{15.7} & \textbf{15.7} & \textbf{15.7} & \textbf{14.0}       & \textbf{13.9}       & \textbf{13.9}           \\
Gaussian Noise  & 43.7       & 43.7       & 43.7       & 37.8       & 37.8       & 37.8           & 25.0          & 25.0          & 25.0          & 12.5       & 12.5       & 12.5             \\
Denoising       & \multicolumn{6}{c}{$H$: 26.2, $W$: 26.2, $S$: 26.2}                           & \multicolumn{6}{c}{$H$: 28.0, $W$: 35.3, $S$: 35.3}                                \\
Restoration     & \multicolumn{6}{c}{$H$: 45.3, $W$: 45.3, $S$: 45.4}                                 & \multicolumn{6}{c}{$H$: 22.0, $W$: 22.0, $S$: 22.0}                               \\ \bottomrule
\end{tabular}
\label{tab:conventional_imagenet}
\end{adjustbox}
\end{table}

\begin{table}[t]
\centering
\caption{PSNR of non-stego (benign) images}
\begin{adjustbox}{width=1.0\linewidth}
\scalebox{1}{
\begin{tabular}{@{}c|cccc|cccc@{}}
\toprule
\multicolumn{1}{l|}{} & \multicolumn{4}{c|}{IMAGENET} & \multicolumn{4}{c}{BOSS1.0.1} \\ \midrule
$\epsilon$               & 1     & 2     & 4     & 8     & 1     & 2     & 4     & 8     \\ \midrule
\textbf{Proposed}     & 50.07 & \textbf{46.30} & \textbf{43.26} & \textbf{39.15} & 55.07 & \textbf{51.30} & \textbf{48.26} & \textbf{44.15} \\
Gaussian Noise        & \textbf{51.24} & 44.13 & 37.80 & 31.77 & \textbf{56.14} & 49.13 & 42.88 & 36.77 \\
Denoising             & \multicolumn{4}{c|}{25.96}    & \multicolumn{4}{c}{31.28}     \\
Restoration           & \multicolumn{4}{c|}{43.12}    & \multicolumn{4}{c}{47.42}     \\ \bottomrule
\end{tabular}
}
\label{tab:orig_imagenet_boss}
\end{adjustbox}
\end{table}

Our method removes the hidden images better than the other methods, as verified by the DC and DT values (at $\epsilon \leq 8$). Compared with the adaptive Gaussian noise method, our method shows an improvement of up to 18$\%$ and 20$\%$ in terms of PSNR, and DC and DT, respectively. By analyzing the DT values, we see that for DS and LSB, hidden information is destroyed even for small $\epsilon$ values. However, for ISGAN, $\epsilon$ should be greater than 2 to see some levels of removal. In addition, we experiment with a conventional steganography technique, LSB insertion. We embedded a full-size gray image into a full-size colored cover image using the LSB insertion technique. 

As shown in Fig.~\ref{fig:result_graph}, we can see that the stego images purified by our method resemble the cover images when increasing the value of $\epsilon$. Also, the degree of removal of the hidden message using our method is similar to or higher than that of the existing steganylsis algorithms. There is no visual degradation of the stego image by applying the proposed method although $\epsilon$ is larger, as shown in Fig.~\ref{fig:change_by_epsilon}. Considering the visual degradation and the destruction of the secret images, $\epsilon = 4$ seems a reasonable value for all cases.

In Fig.~\ref{fig:dsimagenetpsnr}, we compare the visual results when Gaussian noise and PixelSteganalysis are applied against DS as an active warden. Specifically, we compare the degree of destruction of the decoded secret images when the PSNR between the two stego images is almost the same. There is no visual difference between $\hat{x}^{\text{GN}}$ and $\hat{x}^{\text{PS}}$, where GN is Gaussian noise and PS is PixelSteganalysis. However, the results of $\hat{d}$ are completely different. The decoded secret images from $\hat{x}^{\text{GN}}$ are almost preserved, while the decoded secret images from $\hat{x}^{\text{PS}}$ look like noise. As demonstrated in conventional steganography methods~\cite{li2014new,liao2019new} and the background section of our manuscript, the encoding and decoding of secret images are dependent on cover images. If only simple texture regions exist in the cover image and the color is monotonous, the encoding of the secret image cannot be more concentrated in texture regions, so a larger $\epsilon$ is needed to remove the secret image widely embedded all over the cover image. We can see corresponding examples in the 3rd and 6th rows of Fig.~\ref{fig:dsimagenetpsnr}.

The conventional active steganalysis algorithms fail at destroying the hidden secret message. A large amount of alteration on stego images degrades the image quality at the same time. Unlike that, our method rather aims at restoring the original cover image from its stego image. Thus, our method can accurately destroy the hidden secret image with less visual degradation. We also test our method against the recently proposed DL based steganography method, UDH. The results are consistent with the results obtained against DS and ISGAN, and it results are given in supplementary S8. More visual samples for Cifar-10 and Boss1.0.1 datasets are shown in supplementary S11.

\textbf{Analysis about Other Steganalysis Algorithms} An effective active steganalysis algorithm should, first, remove as much of a secret message as possible and, second, do so with minimal degradation of the stego image. That is, both conditions must be met. The Gaussian noise and denoising techniques, however, met the first condition but did not satisfy the second one. The restoration method met the second condition but did not satisfy the first one. In the case of Gaussian noise method, if $\epsilon$ exceeds 4, the degradation of the stego image is severe, and stego image becomes perceptually noisy, as depicted in Fig.~\ref{fig:imagenet_total_output} and the supplementary S10. In the case of denoising, the DT is higher than that of our method sometimes, but the PSNR is very low compared to the other steganography methods. In case of restoration, the PSNR is similar to that of our method; however, DT is typically the worst. The visual samples are presented in supplementary S7.

\begin{figure}[t]
\centering
\includegraphics[width=1.0\linewidth]{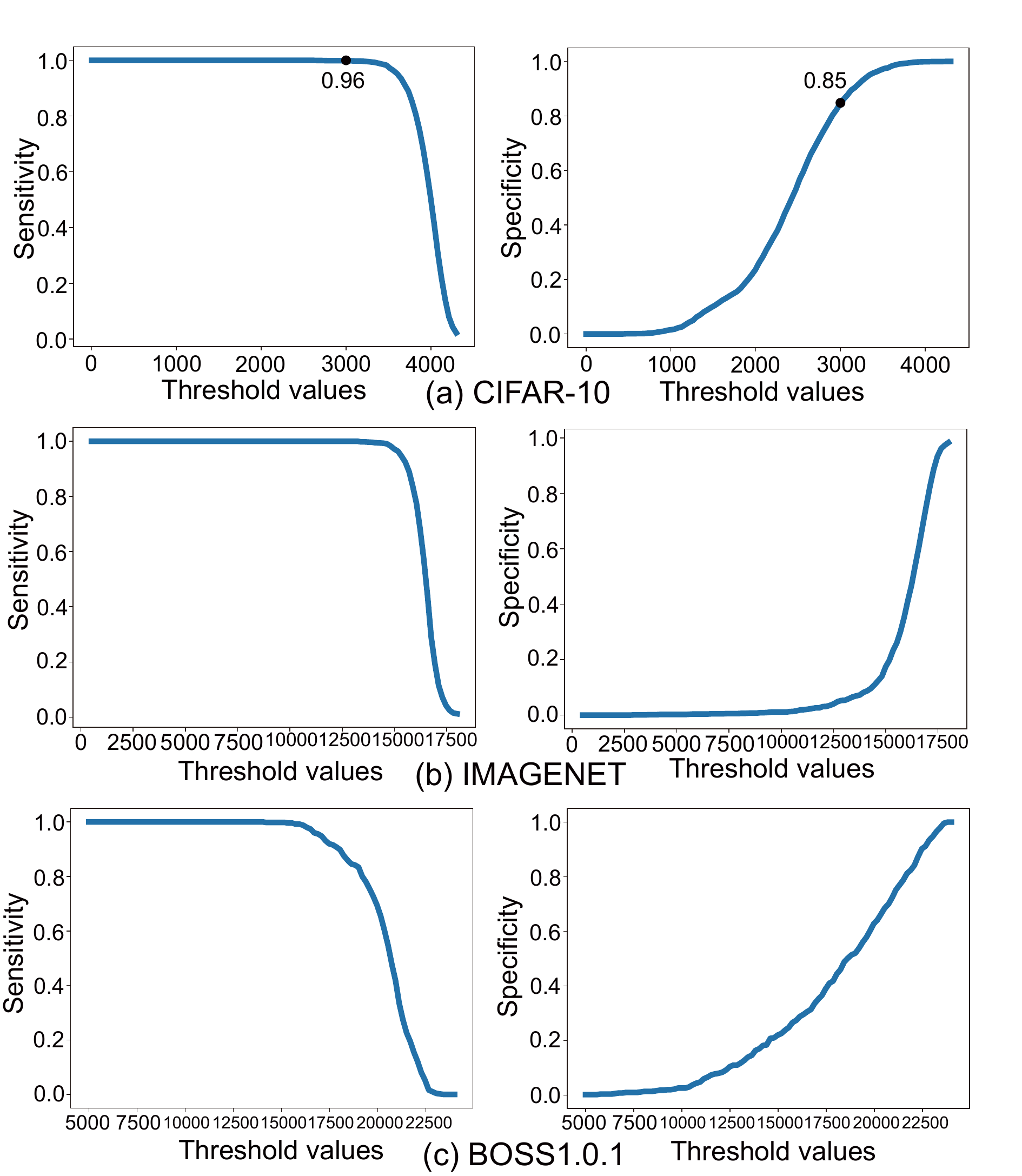}
\caption{Sensitivity and specificity of non-stego and stego (DS) images for the (a) Cifar-10, (b) ImageNet, and (c) Boss1.0.1 datasets. To see the possibility of the proposed method as a passive steganalysis technique, we measured the number of modifications on each non-stego and stego image by the proposed method. The number of modifications on stego images is greater than that of non-stego images, especially for the Cifar-10 and Boss1.0.1 datasets.}
\label{fig:as_detector}
\end{figure}

\textbf{Additional Experiments on Non-DL Steganography} We validate our method not only on DL based steganography algorithms but also on more sophisticated conventional non-DL steganography algorithms (e.g., HUGO, WOW, and S-UNIWARD) by using a small payload (1 bpp/ch) on the ImageNet dataset. Our method is comparable to or even better than other methods on sophisticated conventional non-DL steganography methods in terms of PSNR and DC, as presented in Tab. \ref{tab:conventional_imagenet} (ImageNet) and supplementary S3. Especially, at $\epsilon$ = 2, our method outperforms the other existing methods with an amount of DC decreased by 10 or more on the steganography methods, namely, HUGO, WOW, and S-UNIWARD.

\textbf{Harmless on Non-Stego (Benign) Images} In the case of active steganalysis, we assume having the privilege to allow modifications to all the inspected images. We, therefore, experiment on image degradation when applying our method to non-stego (innocuous/benign) images, as summarized in Tab.~\ref{tab:orig_imagenet_boss} (ImageNet and Boss1.0.1). After applying our proposed method, there was some degree of image degradation, but the PSNR of our method was always higher than that of the other conventional steganalysis algorithms at $\epsilon \leq$ 4. In detail, the standard deviation is less than 1 for all the datasets and $\epsilon$. Our method did not fall below 40 even for the large $\epsilon$ as a whole. Degradation of the Cifar-10 dataset after performing active steganalysis is provided in supplementary S4. In contrast to our proposed method, the existing algorithms mostly show visible degradation of benign images.

\begin{table}[t]
 	\caption{Ablation results of the proposed methods. For PSNR, SSIM and DT, higher is better. For DC, lower is better.}
 	\centering
  	\label{table:abblation_cifar}
  	\begin{adjustbox}{width=1\linewidth}
	    \begin{tabular}{cccccc}
\toprule
                                                                         & $\epsilon$ &  \multicolumn{1}{c}{PSNR~\cite{psnr}} & \multicolumn{1}{c}{SSIM~\cite{ssim}} & \multicolumn{1}{c}{DC~\cite{amritha2016removal}} & \multicolumn{1}{c}{DT} \\ \midrule
                                                                         & 1         & 35.66                  & 0.9834                   & 0.7761                           & 0.2117                            \\
\begin{tabular}[c]{@{}c@{}}Proposed w/o edge \end{tabular} & 2         & 35.72                  & 0.9837                   & 0.7523                           & 0.2365                            \\
                                                                         & 4         & 35.39                  & 0.9811                   & 0.7375                           & 0.2720                            \\ \midrule
                                                                         & 1 & 35.89                  & 0.9839                   & 0.7691                           & 0.2184                            \\
\begin{tabular}[c]{@{}c@{}}Proposed\end{tabular}  & 2 & 35.85                  & 0.9842                   & 0.7258                           & 0.2626                            \\
                                                                         & 4  & 35.67                  & 0.9822                   & 0.6923                           & 0.3001                            \\ \bottomrule
\end{tabular}
	\end{adjustbox}
\end{table}

\begin{figure}[t]
\centering
\includegraphics[width=1\linewidth]{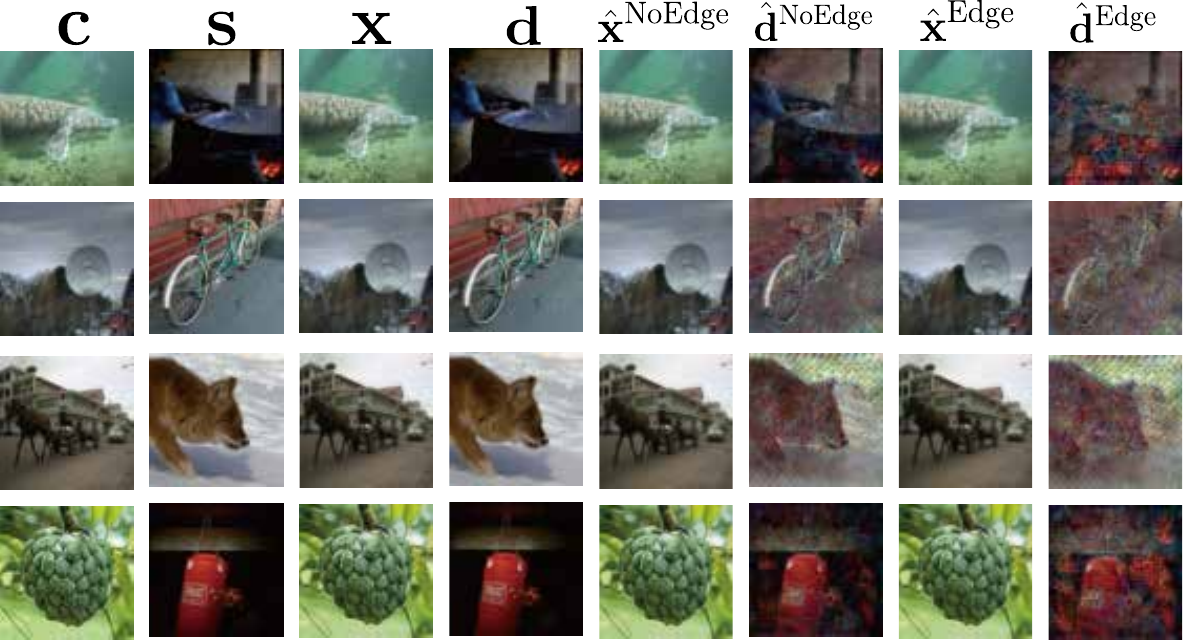}
\caption{Ablation study: no edge detection (DS on ImageNet). $\hat{\mathbf{x}}^{\text{NoEdge}}$ represents a stego image modified by PixelSteganalysis but without edge detection, and $\hat{\mathbf{d}}^{\text{NoEdge}}$ represents a secret image decoded from $\hat{\mathbf{x}}^{\text{NoEdge}}$. $\hat{\mathbf{x}}^{\text{Edge}}$ represents a stego image modified by PixelSteganalysis with edge detection, and $\hat{\mathbf{d}}^{\text{Edge}}$ represents a secret image decoded from $\hat{\mathbf{x}}^{\text{Edge}}$.}
\label{fig:ablation}
\end{figure}

\textbf{Potential as Passive Steganalysis}
If cannot apply our method to every image owing to limited privileges, the method can first operate as a detector (passive steganalysis). Compared to recent advanced passive steganalysis algorithms which assume that access to the cover image, the secret image, or the applied steganography algorithm is required, our algorithm does not need any of them. We measure the total number of modifications of the pixel value on each non-stego and each stego image when applying our method. As a binary classification task, we classify the input image as a stego image when the total number of modifications of the pixel value is greater than or equal to a threshold value, and vice versa. We measure the performance of our method as a detector with various threshold values, as shown in Fig.~\ref{fig:as_detector}. As the metrics, we use the sensitivity and specificity, which is formulated as follows:
\begin{equation}
    \begin{aligned}
        \text{Sensitivity} = \frac{\#\text{True Positive}}{\#\text{True Positive} + \#\text{False Negative}},
    \label{eq:sensitivity}
    \end{aligned}
\end{equation}
\begin{equation}
    \begin{aligned}
        \text{Specificity} = \frac{\#\text{True Negative}}{\#\text{True Negative} + \#\text{False Positive}},
    \label{eq:specificity}
    \end{aligned}
\end{equation}
where a True Positive is a stego image that the model correctly predicts as a stego image, a True Negative is a non-stego image that the model correctly predicts as a non-stego image, a False Positive is a non-stego image that the model incorrectly predicts as a stego image, and a False Negative is a stego image that the model incorrectly predicts as a non-stego image. 

For Cifar-10, the total number of modifications of the pixel value in the stego images was noticeably higher than that of non-stego images. By setting the threshold value to 3,000, it is possible to obtain a sensitivity and specificity of 0.96 and 0.85, respectively. For ImageNet and Boss1.0.1, the total number of modifications of the pixel value in the stego images does not deviate significantly from the total number of modifications of the pixel value in the non-stego images, as compared to the Cifar-10 case.It is because that ImageNet and Boss1.0.1 consist of more various classes of large images. However, as a stego detector, it is possible to obtain a high sensitivity for all three datasets, as shown in Fig.~\ref{fig:as_detector} (a), (b), and (c), with the relatively low threshold values (15,000 for ImageNet and 17,500 for Boss1.0.1). We believe that our method can be used as a primary inspector to find out suspicious images.

\subsection{Ablation Studies} \label{ssec:sec52}

\textbf{No Edge Detection} For ablation studies, we test the effectiveness of edge detection. We proceed the experiment using Cifar-10. We can observe that especially at $\epsilon$ = 1 and 2, the effect of the edge detection as a guide is large. The visual samples of $\textbf{p}_{\mathrm{e}}( \mathbf{x})$ from our neural network are provided in supplementary S9. The quantitative and qualitative results are presented in Tab.~\ref{table:abblation_cifar} and Fig.~\ref{fig:ablation}. As shown in Tab.~\ref{table:abblation_cifar} and Fig.~\ref{fig:ablation}, we could remove more hidden information while keeping the visual degradation of the stego images reduced by making use of edge detection. 

\textbf{No Approximation} The results of Fig.~\ref{fig:result_graph} are obtained using the approximated version. The original version takes an average of 3 min to purify a single Cifar-10 image. However, the approximated version takes less than 10 ms to process the same image. We compare the DC of both the versions by using stego images generated via DS. The results demonstrate that the average DC of the original version is $75.8\%$ and that of the approximated version is $76.3\%$. The difference is as small as $0.5\%$. 

\section{Conclusions}
\label{sec:conclusions}

We propose a DL based steganalysis technique that effectively removes secret images by restoring the distribution of the original image. We use deep neural networks to formulate and solve problems by leveraging sophisticated pixel and edge distributions in the images. In particular, our method shows a remarkable performance against DL based steganography, which is difficult to detect with passive steganalysis in a blind case.

There is a limitation to our method in that it can degrade all of the inspected images. However, when considering an environment or society in which the confidentiality of sensitive information is extremely important, our method will be a good solution. Compared with existing active steganalysis methods, our approach is the only way to restore a stego image to its original cover image. Therefore, our method is feasible for simultaneously minimizing the degradation of benign images and minimizing false negatives. 

In a future study, we will consider replacing $\textbf{p}_{\mathrm{d}}$ with a metric more suited to the characteristics of a steganography algorithm. We currently use the Prewitt operator as $\textbf{p}_{\mathrm{d}}$ to locate the high frequency areas of an image. Moreover, instead of using auto-regressive models, it is possible to use other approaches, such as GAN based methods, to model the distribution of the images. However, the distributions of both the original cover images and the stego images are quite similar; therefore, a significant reduction in the PSNR may occur without a careful modification. Thus, the proposed method can carefully remove suspicious traces at the pixel level by maximally utilizing the intrinsic characteristics of steganography.

\section{Acknowledgments}
\label{sec:ack}

This work was supported by the National Research Foundation of Korea (NRF) grant funded by the Korea government (Ministry of Science and ICT) [2018R1A2B3001628], the BK21 FOUR program of the Education and Research Program for Future ICT Pioneers, Seoul National University in 2021, Institute of Information \& communications Technology Planning \& Evaluation (IITP) grant funded by the Korea government(MSIT) [NO.2021-0-01343, Artificial Intelligence Graduate School Program (Seoul National University)].

\bibliographystyle{IEEEtran}
\bibliography{main.bib}

\begin{thebibliography}{10}
\providecommand{\url}[1]{#1}
\csname url@rmstyle\endcsname
\providecommand{\newblock}{\relax}
\providecommand{\bibinfo}[2]{#2}
\providecommand\BIBentrySTDinterwordspacing{\spaceskip=0pt\relax}
\providecommand\BIBentryALTinterwordstretchfactor{4}
\providecommand\BIBentryALTinterwordspacing{\spaceskip=\fontdimen2\font plus
\BIBentryALTinterwordstretchfactor\fontdimen3\font minus
  \fontdimen4\font\relax}
\providecommand\BIBforeignlanguage[2]{{%
\expandafter\ifx\csname l@#1\endcsname\relax
\typeout{** WARNING: IEEEtran.bst: No hyphenation pattern has been}%
\typeout{** loaded for the language `#1'. Using the pattern for}%
\typeout{** the default language instead.}%
\else
\language=\csname l@#1\endcsname
\fi
#2}}

\bibitem{johnson1998exploring}
N.~F. Johnson and S.~Jajodia, ``Exploring steganography: Seeing the unseen,''
  \emph{Computer}, vol.~31, no.~2, 1998.

\bibitem{sharma2012comparative}
M.~K. Sharma and P.~Gupta, ``A comparative study of steganography and
  watermarking'','' \emph{International Journal of Research in IT \& Management
  (IJRIM)}, vol.~2, no.~2, pp. 2231--4334, 2012.

\bibitem{gardner_2013}
\BIBentryALTinterwordspacing
F.~Gardner, ``How do terrorists communicate?'' Nov 2013. [Online]. Available:
  \url{https://www.bbc.com/news/world-24784756}
\BIBentrySTDinterwordspacing

\bibitem{king2018}
\BIBentryALTinterwordspacing
A.~King, ``Ge engineer tied to china charged with theft of company secrets,''
  Aug 2018. [Online]. Available:
  \url{https://asia.nikkei.com/Business/Companies/GE-engineer-tied-to-China-charged-with-theft-of-company-secrets}
\BIBentrySTDinterwordspacing

\bibitem{lie1999data}
W.-N. Lie and L.~C. Chang, ``Data hiding in images with adaptive numbers of
  least significant bits based on the human visual system,'' in
  \emph{Proceedings 1999 International Conference on Image Processing (Cat.
  99CH36348)}, vol.~1.\hskip 1em plus 0.5em minus 0.4em\relax IEEE, 1999, pp.
  286--290.

\bibitem{pevny2010using}
T.~Pevn{\`y}, T.~Filler, and P.~Bas, ``Using high-dimensional image models to
  perform highly undetectable steganography,'' in \emph{International Workshop
  on Information Hiding}.\hskip 1em plus 0.5em minus 0.4em\relax Springer,
  2010, pp. 161--177.

\bibitem{holub2012designing}
V.~Holub and J.~J. Fridrich, ``Designing steganographic distortion using
  directional filters.'' in \emph{WIFS}, 2012, pp. 234--239.

\bibitem{fridrich2002steganalysis}
J.~Fridrich, M.~Goljan, and D.~Hogea, ``Steganalysis of jpeg images: Breaking
  the f5 algorithm,'' in \emph{International Workshop on Information
  Hiding}.\hskip 1em plus 0.5em minus 0.4em\relax Springer, 2002, pp. 310--323.

\bibitem{dong2018invisible}
S.~Dong, R.~Zhang, and J.~Liu, ``Invisible steganography via generative
  adversarial network,'' \emph{arXiv preprint arXiv:1807.08571}, 2018.

\bibitem{wu2018stegnet}
P.~Wu, Y.~Yang, and X.~Li, ``Stegnet: Mega image steganography capacity with
  deep convolutional network,'' \emph{arXiv preprint arXiv:1806.06357}, 2018.

\bibitem{baluja2017hiding}
S.~Baluja, ``Hiding images in plain sight: Deep steganography,'' in
  \emph{Advances in Neural Information Processing Systems}, 2017, pp.
  2069--2079.

\bibitem{johnson1998steganalysis}
N.~F. Johnson and S.~Jajodia, ``Steganalysis of images created using current
  steganography software,'' in \emph{International Workshop on Information
  Hiding}.\hskip 1em plus 0.5em minus 0.4em\relax Springer, 1998, pp. 273--289.

\bibitem{bachrach2011image}
M.~Bachrach and F.~Y. Shih, ``Image steganography and steganalysis,''
  \emph{Wiley Interdisciplinary Reviews: Computational Statistics}, vol.~3,
  no.~3, pp. 251--259, 2011.

\bibitem{amritha2016removal}
P.~Amritha, M.~Sethumadhavan, and R.~Krishnan, ``On the removal of
  steganographic content from images,'' \emph{Defence Science Journal},
  vol.~66, no.~6, pp. 574--581, 2016.

\bibitem{karaman2012application}
H.~Karaman and S.~Sagiroglu, ``An application based on steganography,'' in
  \emph{Proceedings of the 2012 International Conference on Advances in Social
  Networks Analysis and Mining (ASONAM 2012)}.\hskip 1em plus 0.5em minus
  0.4em\relax IEEE Computer Society, 2012, pp. 839--843.

\bibitem{ettinger1998steganalysis}
J.~M. Ettinger, ``Steganalysis and game equilibria,'' in \emph{International
  Workshop on Information Hiding}.\hskip 1em plus 0.5em minus 0.4em\relax
  Springer, 1998, pp. 319--328.

\bibitem{gou2007noise}
H.~Gou, A.~Swaminathan, and M.~Wu, ``Noise features for image tampering
  detection and steganalysis,'' in \emph{2007 IEEE International Conference on
  Image Processing}, vol.~6.\hskip 1em plus 0.5em minus 0.4em\relax IEEE, 2007,
  pp. VI--97.

\bibitem{shrestha2011general}
P.~L. Shrestha, M.~Hempel, T.~Ma, D.~Peng, and H.~Sharif, ``A general attack
  method for steganography removal using pseudo-cfa re-interpolation,'' in
  \emph{2011 International Conference for Internet Technology and Secured
  Transactions}.\hskip 1em plus 0.5em minus 0.4em\relax IEEE, 2011, pp.
  454--459.

\bibitem{lafferty2008obfuscation}
P.~A. Lafferty, \emph{Obfuscation and the steganographic active warden
  model}.\hskip 1em plus 0.5em minus 0.4em\relax The Catholic University of
  America, 2008.

\bibitem{watermarking}
I.~J. Cox, M.~L. Miller, J.~A. Bloom, and C.~Honsinger, \emph{Digital
  watermarking}.\hskip 1em plus 0.5em minus 0.4em\relax Springer, 2002,
  vol.~53.

\bibitem{imperceptibility}
P.~Ramu, R.~Swaminathan, \emph{et~al.}, ``Imperceptibility—robustness
  tradeoff studies for ecg steganography using continuous ant colony
  optimization,'' \emph{Expert Systems with Applications}, vol.~49, pp.
  123--135, 2016.

\bibitem{mielikainen2006lsb}
J.~Mielikainen, ``Lsb matching revisited,'' \emph{IEEE signal processing
  letters}, vol.~13, no.~5, pp. 285--287, 2006.

\bibitem{filler2010gibbs}
T.~Filler and J.~Fridrich, ``Gibbs construction in steganography,'' \emph{IEEE
  Transactions on Information Forensics and Security}, vol.~5, no.~4, pp.
  705--720, 2010.

\bibitem{furthers-uni}
T.~Denemark, J.~Fridrich, and V.~Holub, ``Further study on the security of
  s-uniward,'' in \emph{Media Watermarking, Security, and Forensics 2014}, vol.
  9028.\hskip 1em plus 0.5em minus 0.4em\relax International Society for Optics
  and Photonics, 2014, p. 902805.

\bibitem{holub2014universal}
V.~Holub, J.~Fridrich, and T.~Denemark, ``Universal distortion function for
  steganography in an arbitrary domain,'' \emph{EURASIP Journal on Information
  Security}, vol. 2014, no.~1, p.~1, 2014.

\bibitem{hayes2017generating}
J.~Hayes and G.~Danezis, ``Generating steganographic images via adversarial
  training,'' in \emph{Advances in Neural Information Processing Systems},
  2017, pp. 1954--1963.

\bibitem{shi2017ssgan}
H.~Shi, J.~Dong, W.~Wang, Y.~Qian, and X.~Zhang, ``Ssgan: Secure steganography
  based on generative adversarial networks,'' in \emph{Pacific Rim Conference
  on Multimedia}.\hskip 1em plus 0.5em minus 0.4em\relax Springer, 2017, pp.
  534--544.

\bibitem{zhu2018hidden}
J.~Zhu, R.~Kaplan, J.~Johnson, and L.~Fei-Fei, ``Hidden: Hiding data with deep
  networks,'' in \emph{Proceedings of the European Conference on Computer
  Vision (ECCV)}, 2018, pp. 657--672.

\bibitem{wang2019stnet}
Z.~Wang, N.~Gao, X.~Wang, J.~Xiang, and G.~Liu, ``Stnet: A style transformation
  network for deep image steganography,'' in \emph{International Conference on
  Neural Information Processing}.\hskip 1em plus 0.5em minus 0.4em\relax
  Springer, 2019, pp. 3--14.

\bibitem{huang2019image}
J.~Huang, S.~Cheng, S.~Lou, and F.~Jiang, ``Image steganography using texture
  features and gans,'' in \emph{2019 International Joint Conference on Neural
  Networks (IJCNN)}.\hskip 1em plus 0.5em minus 0.4em\relax IEEE, 2019, pp.
  1--8.

\bibitem{wang2019hidinggan}
Z.~Wang, N.~Gao, X.~Wang, J.~Xiang, D.~Zha, and L.~Li, ``Hidinggan: High
  capacity information hiding with generative adversarial network,'' in
  \emph{Computer Graphics Forum}, vol.~38, no.~7.\hskip 1em plus 0.5em minus
  0.4em\relax Wiley Online Library, 2019, pp. 393--401.

\bibitem{meng2019novel}
R.~Meng, Z.~Zhou, Q.~Cui, X.~Sun, and C.~Yuan, ``A novel steganography scheme
  combining coverless information hiding and steganography,'' \emph{J. Inf.
  Hiding Privarcy Protection}, vol.~1, no.~1, pp. 43--48, 2019.

\bibitem{chen2020high}
B.~Chen, J.~Wang, Y.~Chen, Z.~Jin, H.~J. Shim, and Y.-Q. Shi, ``High-capacity
  robust image steganography via adversarial network.'' \emph{KSII Transactions
  on Internet \& Information Systems}, vol.~14, no.~1, 2020.

\bibitem{kuppusamy2020novel}
P.~Kuppusamy, K.~Ramya, S.~S. Rani, M.~Sivaram, and V.~Dhasarathan, ``A novel
  approach based on modified cycle generative adversarial networks for image
  steganography,'' \emph{Scalable Computing: Practice and Experience}, vol.~21,
  no.~1, pp. 63--72, 2020.

\bibitem{dlstega1}
A.~Das, J.~S. Wahi, M.~Anand, and Y.~Rana, ``Multi-image steganography using
  deep neural networks,'' \emph{arXiv preprint arXiv:2101.00350}, 2021.

\bibitem{dlstega2}
J.~Qin, J.~Wang, Y.~Tan, H.~Huang, X.~Xiang, and Z.~He, ``Coverless image
  steganography based on generative adversarial network,'' \emph{Mathematics},
  vol.~8, no.~9, p. 1394, 2020.

\bibitem{dlstega3}
J.~Liu, Y.~Ke, Z.~Zhang, Y.~Lei, J.~Li, M.~Zhang, and X.~Yang, ``Recent
  advances of image steganography with generative adversarial networks,''
  \emph{IEEE Access}, vol.~8, pp. 60\,575--60\,597, 2020.

\bibitem{dlstega4}
R.~Meng, S.~G. Rice, J.~Wang, and X.~Sun, ``A fusion steganographic algorithm
  based on faster r-cnn,'' \emph{Computers, Materials \& Continua}, vol.~55,
  no.~1, pp. 1--16, 2018.

\bibitem{yu2018integrated}
C.~Yu, ``Integrated steganography and steganalysis with generative adversarial
  networks,'' 2018.

\bibitem{udh}
C.~Zhang, P.~Benz, A.~Karjauv, G.~Sun, and I.~S. Kweon, ``Udh: Universal deep
  hiding for steganography, watermarking, and light field messaging,''
  \emph{Advances in Neural Information Processing Systems}, vol.~33, pp.
  10\,223--10\,234, 2020.

\bibitem{chaum1988dining}
D.~Chaum, ``The dining cryptographers problem: Unconditional sender and
  recipient untraceability,'' \emph{Journal of cryptology}, vol.~1, no.~1, pp.
  65--75, 1988.

\bibitem{berthold2001web}
O.~Berthold, H.~Federrath, and S.~K{\"o}psell, ``Web mixes: A system for
  anonymous and unobservable internet access,'' in \emph{Designing privacy
  enhancing technologies}.\hskip 1em plus 0.5em minus 0.4em\relax Springer,
  2001, pp. 115--129.

\bibitem{ye2017deep}
J.~Ye, J.~Ni, and Y.~Yi, ``Deep learning hierarchical representations for image
  steganalysis,'' \emph{IEEE Transactions on Information Forensics and
  Security}, vol.~12, no.~11, pp. 2545--2557, 2017.

\bibitem{informationtheory}
K.~Solanki, N.~Jacobsen, U.~Madhow, B.~Manjunath, and S.~Chandrasekaran,
  ``Robust image-adaptive data hiding using erasure and error correction,''
  \emph{IEEE Transactions on image processing}, vol.~13, no.~12, pp.
  1627--1639, 2004.

\bibitem{simmons1984prisoners}
G.~J. Simmons, ``The prisoners’ problem and the subliminal channel,'' in
  \emph{Advances in Cryptology}.\hskip 1em plus 0.5em minus 0.4em\relax
  Springer, 1984, pp. 51--67.

\bibitem{zhu1997algorithm}
C.~Zhu, R.~H. Byrd, P.~Lu, and J.~Nocedal, ``Algorithm 778: L-bfgs-b: Fortran
  subroutines for large-scale bound-constrained optimization,'' \emph{ACM
  Transactions on Mathematical Software (TOMS)}, vol.~23, no.~4, pp. 550--560,
  1997.

\bibitem{goodfellow2014generative}
I.~J. Goodfellow, J.~Pouget-Abadie, M.~Mirza, B.~Xu, D.~Warde-Farley, S.~Ozair,
  A.~Courville, and Y.~Bengio, ``Generative adversarial networks,'' \emph{arXiv
  preprint arXiv:1406.2661}, 2014.

\bibitem{salimans2017pixelcnn++}
T.~Salimans, A.~Karpathy, X.~Chen, and D.~P. Kingma, ``Pixelcnn++: Improving
  the pixelcnn with discretized logistic mixture likelihood and other
  modifications,'' \emph{arXiv preprint arXiv:1701.05517}, 2017.

\bibitem{prewitt1970object}
J.~M. Prewitt, ``Object enhancement and extraction,'' \emph{Picture processing
  and Psychopictorics}, vol.~10, no.~1, pp. 15--19, 1970.

\bibitem{psnr}
Q.~Huynh-Thu and M.~Ghanbari, ``Scope of validity of psnr in image/video
  quality assessment,'' \emph{Electronics letters}, vol.~44, no.~13, pp.
  800--801, 2008.

\bibitem{ssim}
Z.~Wang, A.~C. Bovik, H.~R. Sheikh, and E.~P. Simoncelli, ``Image quality
  assessment: from error visibility to structural similarity,'' \emph{IEEE
  transactions on image processing}, vol.~13, no.~4, pp. 600--612, 2004.

\bibitem{li2014new}
B.~Li, M.~Wang, J.~Huang, and X.~Li, ``A new cost function for spatial image
  steganography,'' in \emph{2014 IEEE International Conference on Image
  Processing (ICIP)}.\hskip 1em plus 0.5em minus 0.4em\relax IEEE, 2014, pp.
  4206--4210.

\bibitem{liao2019new}
X.~Liao, Y.~Yu, B.~Li, Z.~Li, and Z.~Qin, ``A new payload partition strategy in
  color image steganography,'' \emph{IEEE Transactions on Circuits and Systems
  for Video Technology}, vol.~30, no.~3, pp. 685--696, 2019.

\end{thebibliography}

\end{document}